\documentstyle[prl,aps,multicol]{revtex}

\begin{document}
\title{Numerical renormalization group study of the correlation functions
of the antiferromagnetic spin-$\frac{1}{2}$ Heisenberg chain}
\author{}
\author{Karen A. Hallberg}
\address{Max Planck Institut f\"{u}r Festk\"{o}rperforschung,
Heisenbergstr. 1, D-70569 Stuttgart, Germany and \\
Max Planck Institut f\"{u}r Physik komplexer Systeme, Bayreuther Str. 40,
Haus 16, D-01187 Dresden, Germany.}
\author{Peter Horsch}
\address{Max Planck Institut f\"{u}r Festk\"{o}rperforschung,
Heisenbergstr. 1, D-70569 Stuttgart, Germany.}
\author{Gerardo Mart{\'{\i}}nez}
\address{Physics Institute, Federal University of Rio Grande do Sul,
91501-970 Porto Alegre, RS, Brazil
}
\date{\today}
\maketitle

\begin{abstract}

    We use the density-matrix renormalization group technique developed
by White \cite{white} to calculate the spin correlation functions
$<{S}_{n+l}^z{S}_n^z>=(-1)^l \omega(l,N)$ for  isotropic
Heisenberg rings up to $N=70$ sites.
The correlation functions for large $l$ and $N$ are found to obey
the scaling relation $\omega(l,N)=\omega(l,\infty)f_{XY}^{\alpha}
(l/N)$ proposed by Kaplan et al. \cite{horsch} , which is used to
determine $\omega(l,\infty)$.
The asymptotic correlation function $\omega(l,\infty)$ and the
magnetic structure factor $S(q=\pi)$ show logarithmic corrections
consistent with $\omega(l,\infty)\sim a\sqrt{\ln{cl}}/l$, where $c$
is related to the cut-off dependent coupling constant
$g_{eff}(l_0)=1/\ln(cl_0)$, as
predicted by field theoretical treatments.

\end{abstract}


\begin{multicols}{2}

Although the exact ground state of the spin-1/2 chain is explicitly
known, the Bethe-Ansatz wave function is far too complex to derive
directly the spin-correlation functions.
Other methods like bosonization or conformal field theory
have to be used to get information about the asymptotic behavior of
these functions. In general
quantum spin chains and in particular their continuum versions
are very active fields of research, because they serve as a testing
ground of various analytical approaches\cite{frad}.
 Recent field-theoretical studies  predict
the existence of logarithmic corrections to the finite-size scaling of the
energies of these systems
and also to the power law behaviour of the spin correlation
functions stemming from marginally irrelevant operators  .
The modification of the power law originally derived by Luther and
Peschel \cite{luther} has its physical origin in Umklapp scattering
processes which appear in the fermionic representation of the model
after Jordan-Wigner transformation and has been anticipated a few years ago
\cite{haldane}.
Logarithmic corrections to the scaling of the energies and to the
correlation functions  were obtained by Affleck et al. \cite{aff} applying
conformal field theory to Wess-Zumino-Witten non-linear-{$\sigma$} models.
Giamarchi and Schulz \cite{giam} and Singh et al\cite{singh}
used a renormalization technique to study the
sine-Gordon Hamiltonian and obtained to leading order
$(-1)^l (\ln l)^{1/2}/l$ for the
asymptotic decay of the spin correlation function in the case of
the  isotropic Heisenberg chain.

In spite of this analytical progress, numerical attempts have given
contradictory results.
Kubo, Kaplan and Borysowicz\cite{kubo} found a
small logarithmic correction of the form $(\ln l)^{\sigma}/l$ with an
exponent $0.2<\sigma< 0.3 $ instead of $0.5$ as predicted by theory.
Later  attempts to check  the theoretical prediction were made by
Liang\cite{liang} and by Lin and Campbell\cite{lin1}, who reported the
absence of logarithmic corrections ($\sigma \cong 0$)
for spin-1/2 chains, whereas Sandvik and Scalapino\cite{sandvik} report
an exponent $\sigma\simeq 1/2$.

Progress on the numerical side was hampered mainly because highly
accurate diagonalization results could be obtained only up to $N=30$,
while Monte Carlo data for larger systems had too large statistical errors.
An equally important reason was that it had been assumed that the data can
be analyzed using the universal asymptotic law $(\ln l)^{\sigma}/l$
\cite{kubo,liang,lin1,sandvik}. In this paper we clarify and conclude
this long dispute on the numerical evidence of the logarithmic corrections
by considering the nonuniversal scaling of the coupling constant.

We use the density matrix algorithm (DMA) \cite{white} to study
the large-distance decay of the
correlation functions and find that they can be calculated
 with such high precision
for sufficiently large systems that the
subtle logarithmic corrections to the correlation functions can be
resolved.
This technique leads to highly accurate results for much larger systems than
those which can be solved by straightforward exact diagonalization.
The DMA allows for a systematic truncation of the Hilbert space
by keeping the most relevant states in describing a state ({\it e.g.}
the ground state)
of a larger system, instead of the lowest energy states usualy kept in
previous real space renormalization techniques.
A general iteration of the method consists of:
i)The effective Hamiltonian defined for the superblock 1+2+1'+2'
(where the blocks 1 and 1'
come from previous iterations and blocks 2 and 2' are new added ones) is
diagonalized to obtain the ground state $|\psi>$ (other states could be also
kept).
ii) The density matrix $\rho_{ii'}=\sum_j \psi_{ij} \psi_{i'j}$ is constructed,
where $\psi_{ij}=<i\otimes j|\psi>$, the states $|i>$ ($|j>$) belonging to the
Hilbert space of blocks 1 and 2 (1' and 2'). The eigenstates of
$\rho$ with the  highest
eigenvalues (equivalent to the most probable states of blocks
1+2 in the ground state of the superblock) are kept up to a certain cutoff.
iii) These  states form a new reduced basis to which all the operators
have to be changed and the block 1+2 is renamed as block 1.
iv) A new block 2 is added (one site in our case) and the new superblock
(1+2+1'+2') is formed as the direct product of the states of all the blocks
(the blocks 1' and 2' are identical to blocks 1 and 2 respectively).

The method has been applied successfully to several problems
such as the Haldane gap of spin-1 chains \cite{spin1}, the one-dimensional
Kondo-insulator \cite{kondo} and the two-chain Hubbard model \cite{twochain}.

    We used the DMA method
keeping up to
200 states per block, one target state (the ground state) and periodic
 boundary
conditions. The  ground state energy
has a relative error of $\sim 10^{-6}$ for a system with $N=70$ sites and
$\sim 10^{-5}$ for $N=100$, as compared to the exact
finite-size energies calculated
using the Bethe-Ansatz. The error in the spin correlation function for
large distances
 is almost
two orders of magnitude larger than the error in the energy, as we estimated
by comparing our calculations with
exact results for 30 sites \cite{lin1}. Because of this we present
results up to $N=70$ so as to have a large enough accuracy.

In the following we consider the average value
${\bar{\omega}}(l,N)={1\over 4}[\omega (l-1,N)+2\omega (l,N)+
\omega (l+1,N)]$
(where $<{S}_{n+l}^z{S}_n^z>=(-1)^l \omega(l,N)$)
to remove  the even-odd-$l$ oscillations of the spin correlation
function, which are of order $l^{-2}$ in $\omega$  and
$\sim l^{-4}$ in $\bar{\omega}$ \cite{aff3}.
    To extract the values of the correlation function for
the infinite system, we adopt the scaling relation of Kaplan et al.
\cite{horsch}
\begin{equation}
\label{eq:one}
{\bar{\omega}}(l,N)=\bar\omega(l,\infty) f(l/N),
\end{equation}
which is expected to hold for sufficiently large $l$ and $N$. The
scaling function is given by $f(l/N)={f_{XY}}^{\alpha}(l/N)$ in terms of
the scaling function of the XY-model
\begin{equation}
f_{XY}(x)=1+.28822 \sinh^2(1.673 x).
\end{equation}

To check the scaling we plot in Fig. 1
\begin{equation}
Z(l,N)={\bar{\omega}}(l,N)/{\bar{\omega}}(l,2l)
\end{equation}
as a function of $l/N$.
We find that all the curves coincide within 0.3\%
 in a unique curve for $l\geq 7$, which confirms the scaling hypothesis.
 Following the proposal of Ref.~\cite{horsch} we performed a least square fit
for  these curves
using the function
\begin{equation}
Z(l,N)=\left[ {f_{XY}(l/N)\over f_{XY}(1/2)}\right ]^{\alpha}
\end{equation}
and obtained for the exponent
$\alpha =1.805_2$ (the small error is determined from the
dispersion of the fitted exponents considering all the curves).
The fit is also shown in Fig. 1.
 This implies that the scaling function defined in Eq. 1 is:
\begin{equation}
\label{eq:f}
f(x)=\left[ 1+.28822 \sinh^2(1.673 x)\right ]^{1.805}
\end{equation}

The surprising quality of the scaling is shown in Fig.~2 where we plot
the averaged ${\bar{\omega}}(l,N)$ as a function of $l/N$ for several values
of $l$.
 From Eq.1 we see that the size dependence of the correlation
function is given by the function $f(l/N)$ rather than the
normally used $1/N^2$ behaviour.
We
have fitted these curves with a one-parameter fit using
Eq.~\ref{eq:one}  with $\bar{\omega}(l,\infty)$ as the free parameter.
We use these extrapolated values  $\bar{\omega}_e (l,\infty)$
to study the logarithmic corrections below.

There are very small deviations from scaling
which cannot be seen in Fig.2 , which look systematic and are
probably not due to the Hilbert-space truncation.
They appear as a small $N$-dependence of
${\bar{\omega}}(l,N)/f(l/N)$. These deviations are less than
0.4\% of the value at $l/N=1/2$.
We note that these deviations could
induce an error on $\bar{\omega}(l,\infty)$
 of at most $1\%$
for the largest $l$-values.

To display the logarithmic behaviour of the correlation function we have
plotted  $l{\bar{\omega}}(l,N)/f(l/N)$
as a function of $l$ ($l\geq 7$) for different $N$ values in Fig. 3.
Here we can see the small $N$-dependence for a given $l$ mentioned above, but
a clear logarithmic behaviour is found.  We also show the extrapolated
values
$l \bar{\omega}_e (l,\infty)$.
To visualize the magnitude of the finite-size
correction to the correlation functions,
we show in the inset the bare data for a finite system,
$l{\bar{\omega}(l,N)}$ as compared to the finite-size corrected
values $l{\bar{\omega}(l,N)}/f(l/N)$.
We see that the corrections
are largest for $l\simeq N/2$ (as can also be  seen from Eq.~\ref{eq:f})

The logarithmic corrections to the correlation function follow from
the scaling relation\cite{aff,giam,singh}
\begin{equation}
\frac{G_i(r)}{G_i(r_0)} =\frac{r_0}{r}\exp\left( -4\pi b_i \int^r_{r_0}
d(\ln{r^{\prime}}) g(r^{\prime})\right),
\end{equation}
where we use the notation of Ref. 6 with $G_i(r)\equiv
{\bar\omega}(r,\infty)$
and $l$ replaced by the continuous variable $r$.
The $r$-dependence of the coupling constant to one-loop order is given
as
\begin{equation}
g(r)=\frac{g(r_0)}{1 + \pi b g(r_0) \ln(r/r_0)}.
\end{equation}
For spin-1/2 the parameters are determined by $b=4/\sqrt{3}$ and
$4b_i/b =-1/2$. The $r$-dependence of the coupling constant leads
after integration to the multiplicative logarithmic correction of the
form
\begin{equation}
\frac{G_i(r)}{G_i(r_0)} =\frac{r_0}{r}
\sqrt{\frac{g(r_0)}{g(r)}}.
\end{equation}
For the following analysis we insert Eq.(7) and obtain
\begin{equation}
\frac{G_i(r)}{G_i(r_0)} =\sqrt{g_{eff}} \frac{r_0}{r}
\left[ \ln(\frac{r}{r_0} e^{1/g_{eff}})\right]^{\frac{1}{2}} ,
\end{equation}
with $g_{eff}(r_0)\equiv 4\pi g(r_0)/\sqrt{3}$. This equation shows
that the universal, i.e. coupling constant independent, asymptotic
relation $\frac{1}{r}(\ln r)^{1/2}$ is only reached for sufficiently
large distances $r$.

We fitted the curves in Fig. 3 with the dependence predicted by
renormalization group
\begin{equation}
\label{eq:log}
l{\bar{\omega}}(l,\infty)=a\sqrt{\ln(cl)} ,
\end{equation}
where $c=\exp(1/g_{eff}(l_0))/l_0$
defines the scale on which the asymptotic behavior
$\omega(l,\infty)\sim \sqrt{\ln(l)}/l$
is approached. Figure 4 gives a comparison of the data and our best fit with
$a=6.789\times 10^{-2}$ and $c=23.21$ . This implies that the asymptotic
regime, i.e. $\ln(c)/\ln(l) << 1$,
is reached only for chains with more than several thousand sites.
The small deviations may either stem from remaining uncertainities
in the determination of $\bar{\omega}_e(l,\infty)$ or from the
one-loop calculation, which is exact only to order ${\cal O}(g^2)$.
 From the value for $c$ we obtain for the coupling constant $g_{eff}(20)=
0.163$ at $l_0=20$. This may be compared with an effective coupling constant
$g_{eff}(20)\sim 0.26$ deduced numerically from the scaling of the
ground state energy and triplet and singlet excitations \cite{aff}.

We stress that our analysis of the data differs from earlier numerical
studies where the asymptotic expression  $ {\bar{\omega}}(l,\infty)
\cong a (\ln l)^{\sigma}/l$ was considered assuming $c=1$ and with
$\sigma$ as a free parameter\cite{kubo,liang,lin1}.
Sandvik and Scalapino\cite{sandvik} on the other hand
suggested that previous numerical
studies\cite{kubo,liang,lin1} did not succeed in finding the proper
exponent since the scaling relation (1) may not hold in presence of
logarithmic corrections. They proposed instead an alternative relation
which connects $\omega(l,N)$ with the asymptotic correlation function
$\omega(l,\infty)$ and which does not obey (1)\cite{17}.
After subtracting the oscillatory $1/l^2$-contribution from the
correlation function they analyse the ratio
$D(l,N)=\omega(l,N) l/\sqrt{\ln{l}}$, assuming the log-corrections are
given by $\sqrt{\ln{l}}$. Given our results $D(l,N)\propto
f(l/N)\sqrt{\ln{cl}}/\sqrt{\ln{l}}$, i.e. the functional used in
Ref.12 accounts for finite size effects but also for parts of the
log-corrections, consequently their analysis concerning the form of
these corrections is not conclusive.

Finally it should also be possible to determine the logarithmic
corrections from the $N$-dependence of the structure factor
$S_N(q=\pi)=\sum_l {\bar{\omega}}(l,N)$ .
Earlier attempts\cite{kubo,liang,lin1,karb} were not successful
because this expression also involves large finite size effects
as $ {\bar{\omega}}(l,N) \cong  {\bar{\omega}}(l,\infty)f(l/N)$.
It is therefore better to consider the quantity
\begin{equation}
\label{eq:S}
S_N(\pi)=\sum^{N/2}_{-N/2 +1}\frac{{\bar{\omega}}(l,N)}{f(l/N)} .
\end{equation}
Given Eq.~\ref{eq:log} for $ {\bar{\omega}}(l,\infty)$ at large $l$ one
expects
$S_N(\pi)\cong const+\frac{4}{3} a \ln^{3/2}(cN/2)$. From our fit
of $S_N$
we find values for $a$ and $c$ which are very close
to the parameters deduced from $l {\bar{\omega}}(l,\infty)$ (see Fig. 5).

    In conclusion we have shown that the scaling relation proposed in
\cite{horsch} provides a very accurate description of finite-size effects
 for large enough systems
and distances $(l\geq 7)$,
 and we have obtained an improved value for
the exponent entering the expression for the scaling function.
Furthermore we have shown that the correlation function $\omega(l,\infty)$
of the infinite system, which can be determined from $N$-site rings
for $l\leq N/2$, does not obey the universal asymptotic law
$(\ln l)^{\sigma}/l$ as was assumed in previous numerical work, but is
governed by the nonuniversal scaling of the coupling constant.
Our data confirms the multiplicative logarithmic corrections to the
spin correlation function of the form  $\omega(l,\infty)\sim
a\sqrt{\ln{cl}}/l$ as derived from quantum field theory
\cite{aff,giam,singh}, and we
determine in particular the scale parameter $c$ of the logarithmic
term and the related effective coupling constant
which has not been obtained by these approaches.

We acknowledge useful discussions with T. Giamarchi, H. Schulz, H. Eskes
and R. Zeyher and also with X. Wang concerning the numerical technique.
G. M. gratefully acknowledges support from the Max-Planck
Institutes FKF (Stuttgart) and PKS (Dresden) during his stay.
{}~\\
{}~\\
{}~\\
{}~\\
{}~\\

\narrowtext
\begin{figure}
\caption{The ratio $Z(l,N)={\bar{\omega}}(l,N)/{\bar{\omega}}(l,2l)$
versus $l/N$ for $l\geq 7$ and $ 14\leq N \leq 70$. The scaling
function (4) with $\alpha=1.805$ is shown as solid line.
\label{fig:one}}
\end{figure}

\begin{figure}
\caption{Correlation function $\bar{\omega}(l,N)$ versus $l/N$
for $l=7-29$. The solid curves are single parameter fits using the
scaling relation (1) with $\bar{\omega}(l,\infty)$ as free parameter.
\label{fig:two}}
\end{figure}

\begin{figure}
\caption{Comparison of $l {\bar{\omega}}(l,N)/f(l/N)$ versus $l$ for $N=14-70$
with $l {\bar{\omega}}_e(l,\infty)$ (solid curve).
  In the inset $l \bar{\omega}(l,N)/f(l/N)$ (filled circles)
 and $l \bar{\omega}(l,N)$ (open circles)
are shown for $N=70$ showing how $f(l/N)$ corrects the finite-size effects.
\label{fig:three}}
\end{figure}

\begin{figure}
\caption{Logarithmic corrections are clearly seen in the quantity
$l {\bar{\omega}}_e(l,\infty)$ versus $l$, which is well fitted by the
analytical
expression (10) (solid curve).
\label{fig:four}}

\end{figure}

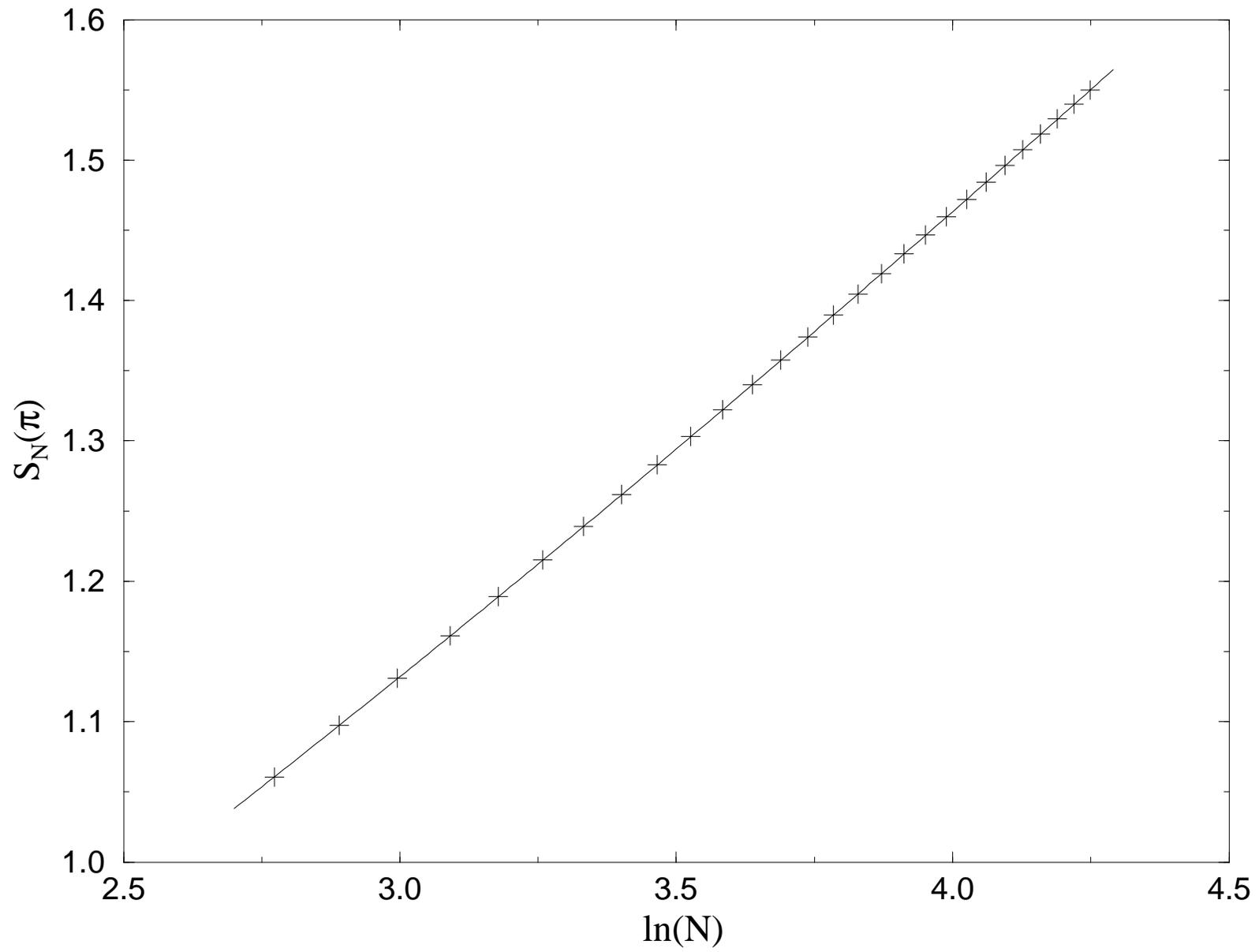
\begin{figure}
\caption{$S_N(\pi)$ (Eq. 11) vs. $\ln (N)$. Comparison of data ($+$) and fit
(solid line) using $S_N(\pi)
\protect\cong const+\frac{4}{3} a \ln^{3/2}(cN/2)$ with $const=
-4.06\;10^{-2}$, $a=6.67\;10^{-2}$ and $c=25.5$.
\label{fig:five}}
\end{figure}

\end{multicols}

\begin{references}
\bibitem{17} If we consider $Z(l/N)$, the largest deviations in the data
(Fig. 1) around $l/N\sim 0.2$ are $\pm 0.2\%$; however, when we calculate
this function from Sandvik and Scalapino's expression the deviations are
$\sim 2\%$, {\it i.e.} an order of magnitude larger.
\end{references}
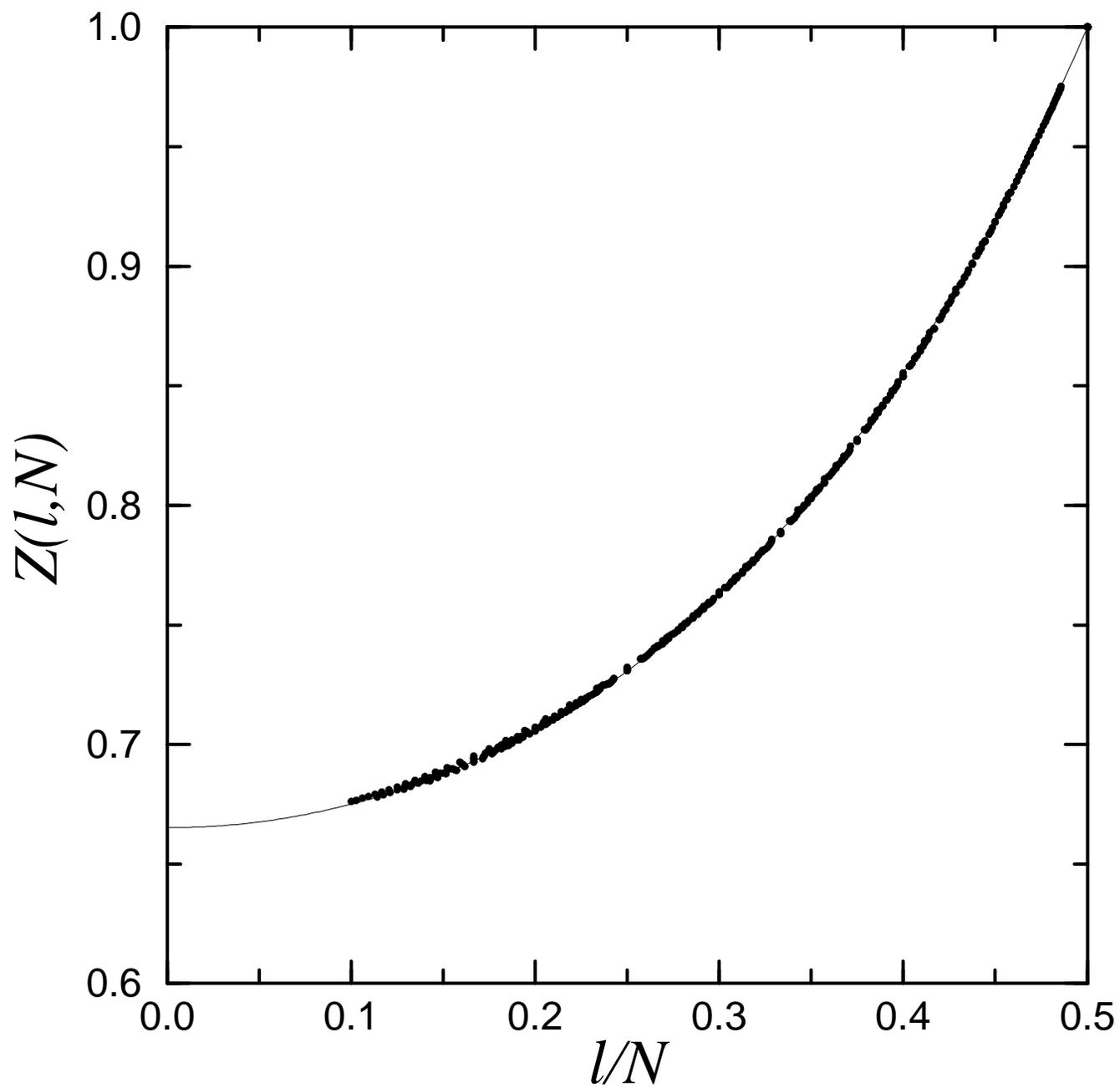
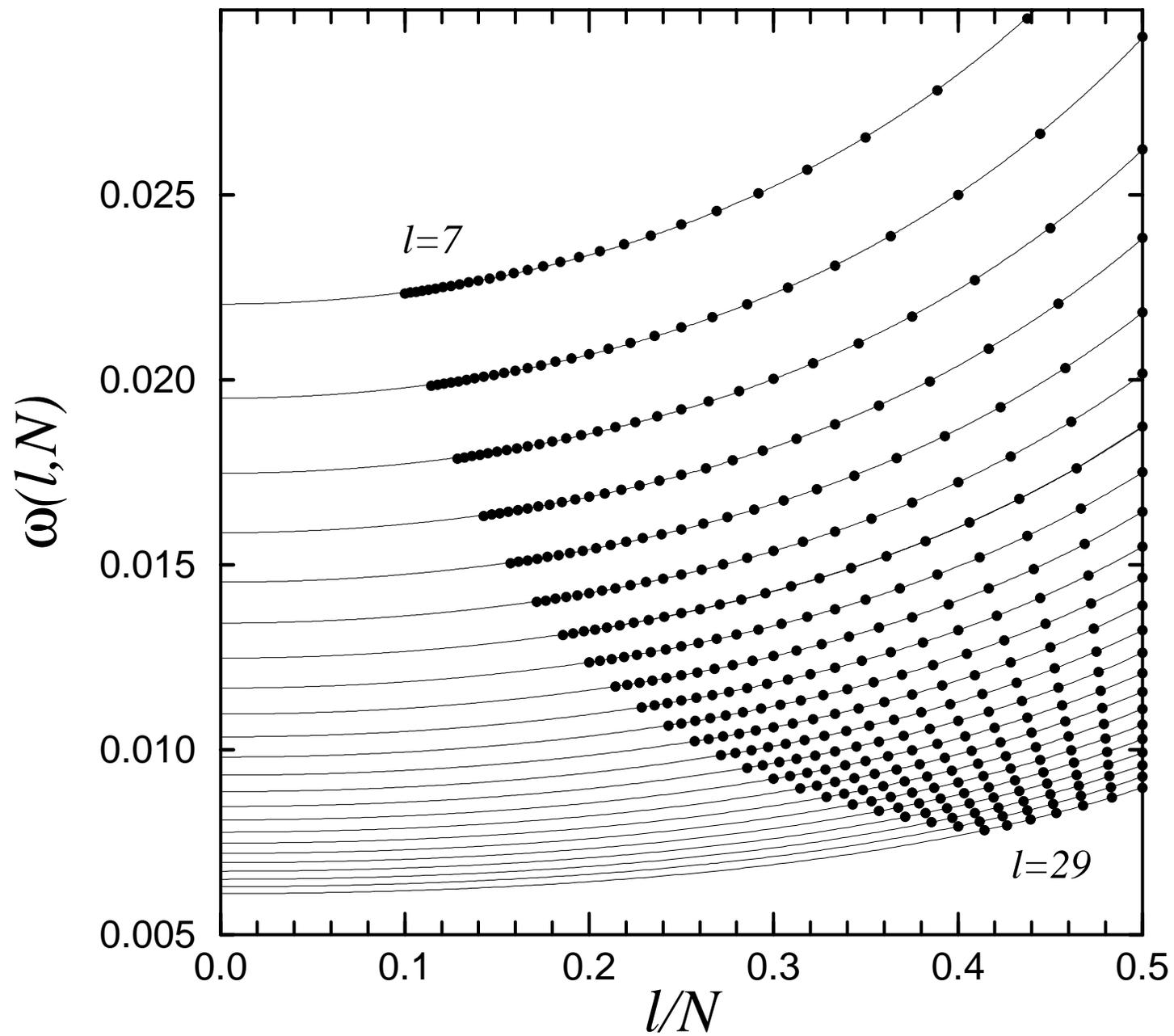
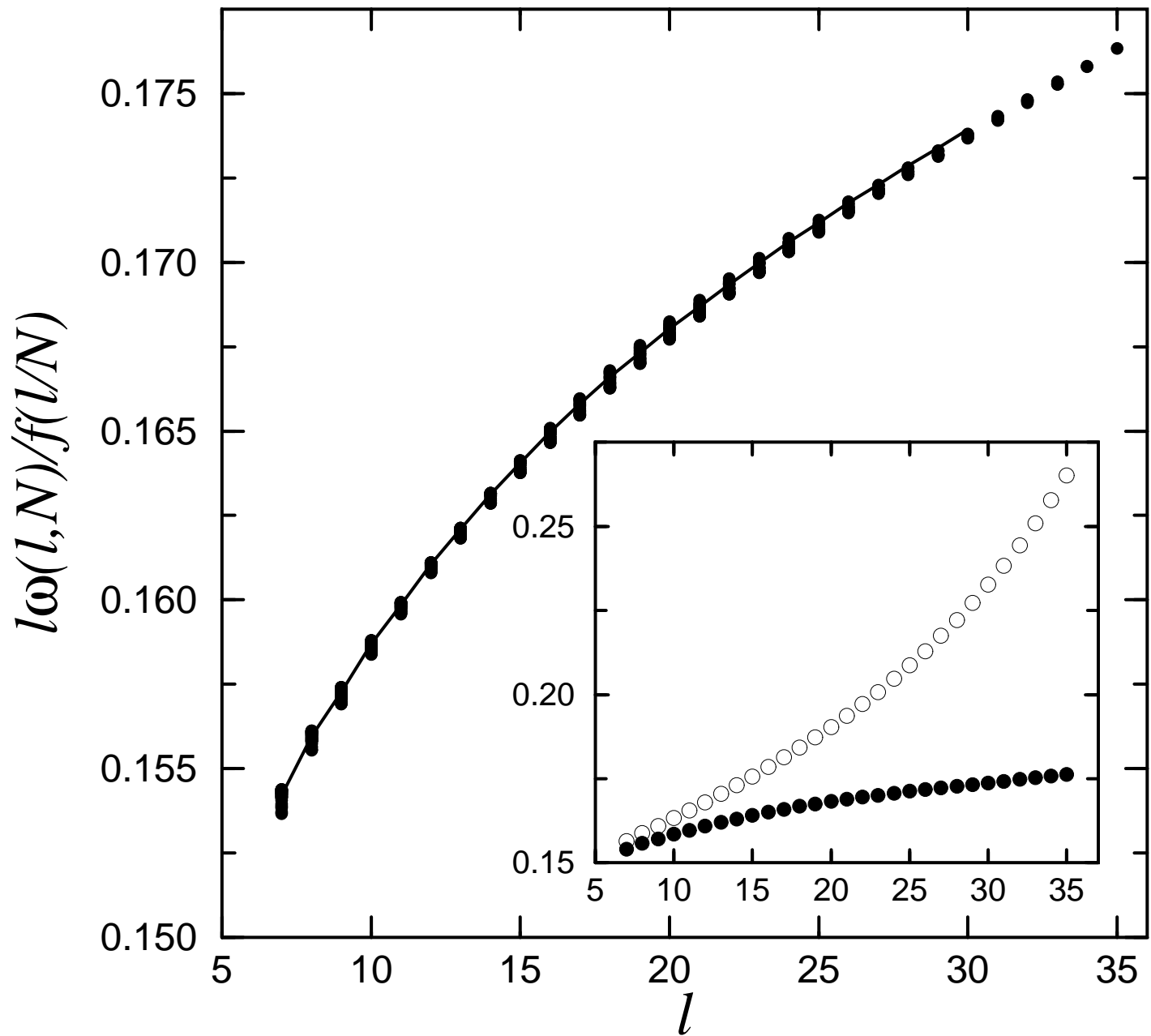
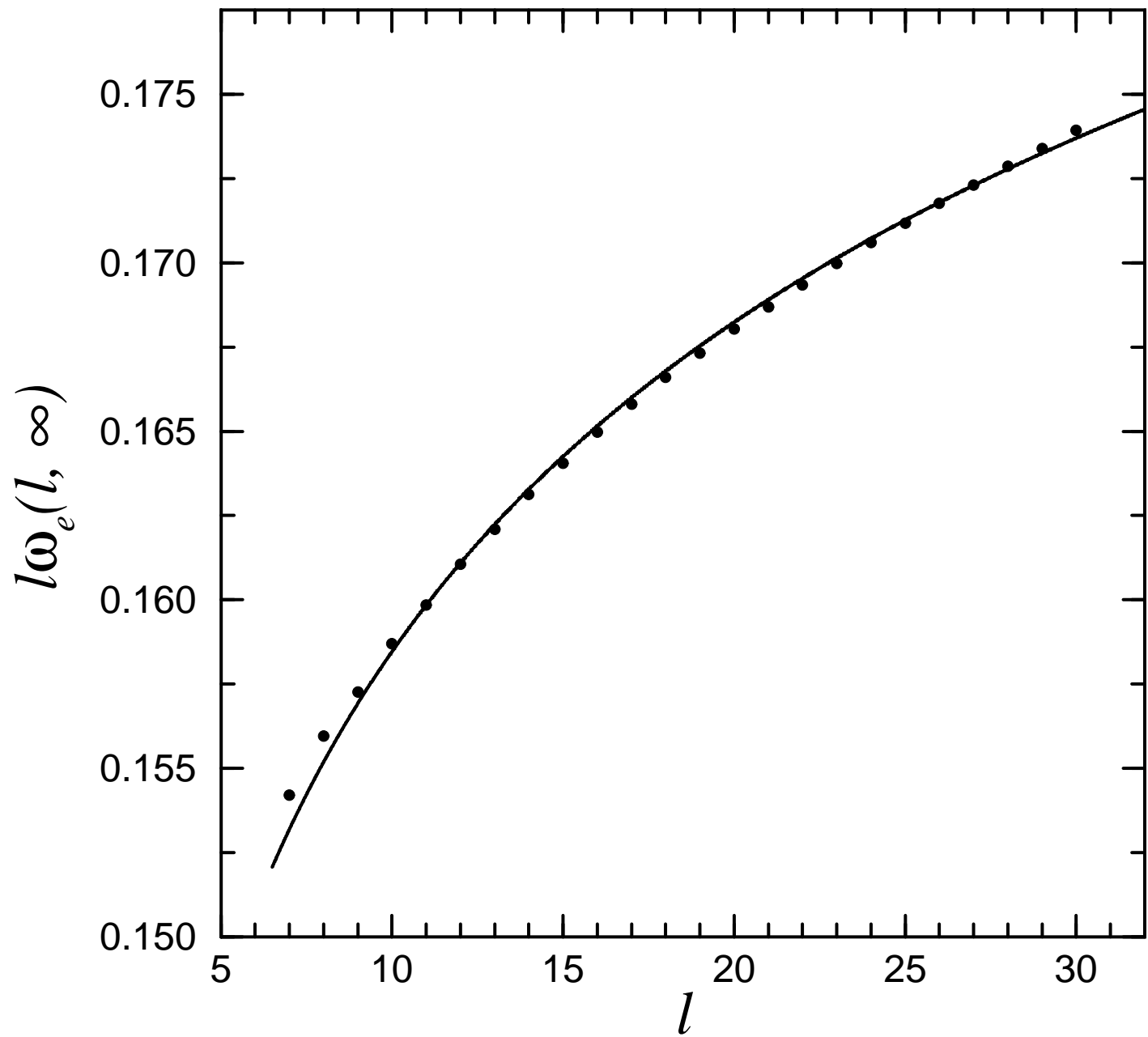
\end{document}